\documentclass[apjl]{emulateapj}

\newcommand{\ctbd}[1]{}

\newcommand{\cfa}{Harvard-Smithsonian Center for Astrophysics (CfA)}

\newcommand{\Ks}{\ensuremath{\rm K_S}}
\newcommand{\masy}{\ensuremath{\rm mas\,yr^{-1}}}
\newcommand{\kms}{\ensuremath{\rm km\,s^{-1}}}

\newcommand{\teff}{\ensuremath{T_{\rm eff}}}
\newcommand{\logg}{\ensuremath{\log{g}}}
\newcommand{\vsini}{\ensuremath{v \sin{i}}}

\newcommand{\msun}{\ensuremath{M_\sun}}

\newcommand{\mjup}{\ensuremath{M_{\rm J}}}

\newcommand{\msini}{\ensuremath{m \sin i}}

\newcommand{\prog}[1]{\textsc{\lowercase{#1}}}

\newcommand{\fihat}{\prog{fihat}}
\newcommand{\fistar}{\prog{fistar}}

\newcommand{\figr}[1]{Fig.~\ref{fig:#1}}
\newcommand{\secr}[1]{\S\ref{sec:#1}}

\defcitealias{bouchy05}{B05}

\newcommand{\hds}{\mbox{HD 189733}}
\newcommand{\hdsb}{\mbox{HD 189733b}}
\newcommand{\hdsB}{\mbox{HD 189733B}}

\newcommand{\hdsBm}{\mbox{2MASS 20004297+2242342}}

\shorttitle{A stellar companion in the \hds\ system}
\shortauthors{Bakos et al.}

\begin{document}
\title{A stellar companion in the HD 189733 system with a known
transiting extrasolar planet}

\author{
G\'asp\'ar~\'A.~Bakos\altaffilmark{1,2}, 
Andr\'as~P\'al\altaffilmark{3,1},
David~W.~Latham\altaffilmark{1},
Robert~W.~Noyes\altaffilmark{1},
Robert~P.~Stefanik\altaffilmark{1}
}
\email{gbakos@cfa.harvard.edu}

\altaffiltext{1}{\cfa,
	60 Garden Street, Cambridge, MA 02138, USA}
\altaffiltext{2}{Hubble Fellow}
\altaffiltext{3}{E\"otv\"os Lor\'and University, Department of
	Astronomy, H-1518 Budapest, Pf.~32., Hungary}

\begin{abstract}
We show that the very close-by (19 pc) K0 star HD 189733, already found
to be orbited by a transiting giant planet, is the primary of a
double-star system, with the secondary being a mid-M dwarf with
projected separation of about 216 AU from the primary. This conclusion
is based on astrometry, proper motion and radial velocity measurements,
spectral type determination and photometry. We also detect differential
proper motion of the secondary. The data appear consistent with the
secondary orbiting the primary in a clockwise orbit, lying nearly in
the plane of the sky (that is, nearly perpendicular to the orbital
plane of the transiting planet), and with period about 3200 years.
\end{abstract}


\section{Introduction}
\label{sec:intro}


Of the $\sim 170$ exoplanets in 146 planetary systems known at the
present
time\footnote{http://vo.obspm.fr/exoplanetes/encyclo/index.php},
the majority ($\sim160$) are revealed only by the reflex
velocity of their parent stars, \citep[e.g.~][]{mayor95,marcy96},
yielding their periods, eccentricities, semi-major axes, and \msini\ 
values. However, a small number (9) of the known exoplanets transit
their host stars, so that the inclination ambiguity is removed and \---
assuming a mass and radius for the primary \--- 
their actual mass and radius, hence their mean density may be determined.

A number of the known exoplanets have been found to occur in
multiple stellar systems.
The small sample of 22 
such systems form an important class of objects 
\citep[for references, see: ][]{eggenberger04b,eggenberger05a,
mugrauer04a,mugrauer04b,mugrauer05a}.
Until now, however, no planet found in a multiple stellar system also
transits its parent star.  In this Letter, we note that the parent star 
of the recently-discovered transiting extra-solar planet
\hdsb\ \citep[][hereafter \citetalias{bouchy05}]{bouchy05}
is itself a member of a double-star system.  

In \secr{comp} we present the evidence that the star \hds\ has a
physical companion star.  
Specifically, 
\secr{comprop} shows that it has
a common proper-motion companion;
\secr{radvel}
shows that the companion's radial velocity is the same as
\hds\ within uncertainties; and \secr{char} shows that the companion is a
red dwarf star which must lie at approximately the same distance as
\hds. For these combined reasons the co-moving companion, now labeled
\hdsB, almost certainly must be a true physical companion. In \secr{orb} we 
detect a differential proper motion, and 
obtain an initial estimate of the orbital motion of \hdsB\ about \hds.
Finally, \secr{conc} discusses some implications of this
finding, and avenues for future work.

\section{Evidence that \hds\ has a Physical Companion Star}
\label{sec:comp}


\hds\ is a close-by ($D=19.3\pm0.3$pc) K0 dwarf, with mass of
$0.82\pm0.03\msun$ and other properties as described in
\citetalias{bouchy05}. The 2MASS survey
\citep{cutri03} lists a nearby red star, \hdsBm\ 
($\rm J = 10.12\pm0.04$, $\rm H = 9.55\pm0.08$, $\Ks = 9.32\pm0.03$,
$\rm J-K_S=0.8$),
some 3.7 magnitudes fainter in \Ks\ and 5 magnitudes fainter in V
(estimated from J, H, \Ks).  This star has an angular separation of
11.2\arcsec\ from \hds, lying 10.2\arcsec\ W and 4.6\arcsec\ to the S.

\hds\ is not listed in the Washington Double Star Catalogue
\citep[WDS;][]{mason01}, but here we will show that \hdsBm\ is in fact
its physical companion; in anticipation of that result we henceforth
denote the companion as \hdsB. Note that the latter name is distinct
from \hdsb, the name given by \citetalias{bouchy05} to the planetary
companion to \hds. Our conclusion that the two stars form a bound
binary system is based on the following evidence.

\subsection{Common Proper Motion}
\label{sec:comprop}

\hds\ has a relatively high proper motion of 
$\mu_{\alpha} = -2.49\pm0.68$ \masy\ and 
$\mu_{\delta} = -250.81\pm0.53$ \masy 
\citep[Hipparcos; ][]{perryman97}. 
To determine whether \hds\ and \hdsB\ share common proper
motion, and hence may be members of the same physical system, we have
inspected the following archival material:
i) the digitized Palomar Observatory Sky Survey (POSS)
scans\footnote{http://archive.stsci.edu/cgi-bin/dss\_form}
(POSS-I, 1951 R-band; POSS-II, 1990 R-band, 1992 B-band, and 1996
I-band); 
ii) the HST QuickV survey (1982, R-band),
iii) the 2MASS\footnote{http://irsa.ipac.caltech.edu/} 2000 J-, H- and
K-band 
scans \citep{skrutskie00}.
In addition, in November 2005, we used the TopHAT telescope of the HAT
Network at the Fred Lawrence Whipple Observatory (FLWO), to obtain
I-band images.
We also acquired 8 short exposure
I-band frames in December 2005 with KeplerCam on the 1.2m telescope at FLWO.

\notetoeditor{This is the original location where Fig.1 was placed in
the latex file. If possible, please place Fig.1 om top of page 2}

\begin{figure*}[t]
\plotone{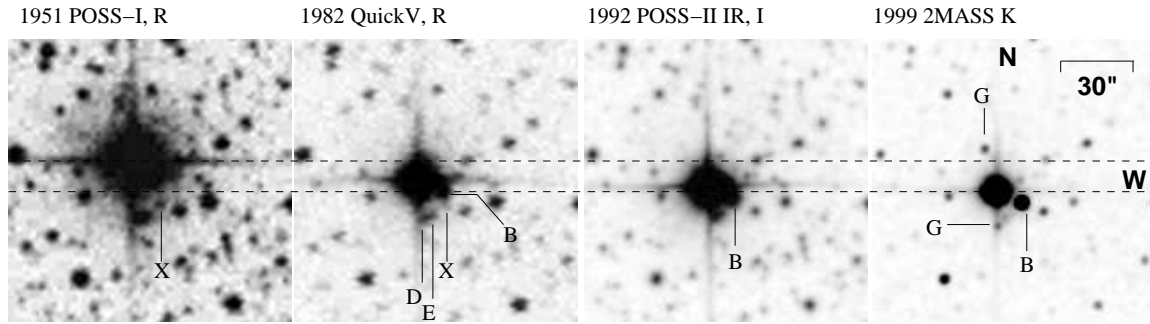}
\caption{
\hds\ in four different epochs on registered frames. The
known proper motion to the South is well visible. The upper dashed line
shows the declination in 1951, and the lower dashed line for 1999. The
companion \hdsB\ is not visible on the POSS-I frame (left panel) because
\hds\ is over-saturated and the scan resolution is not
adequate. \hdsB\ (indicated by {\bf B}) is visible on the rest of the
panels, displaced to the SW by $\sim 11\arcsec$ from \hds; it is seen
increasingly better from left to right, as the companion is
relatively brighter at increasingly longer wavelengths (R, I and K
bands). ``X'' denotes a faint star visible on POSS-I and the QuickV
scans, before the Southward moving \hdsB\ merges with it.
``D'' and ``E'' mark two
faint stars that are visible on all of the frames, but one of the
artificial filter glints (``G'') merges with their position on the
2MASS frame.
\label{fig:hd189mon}}
\end{figure*}

We used our home-grown \fihat\ software environment (P\'al et al.~2006,
in preparation) to find sources on all the above images ($\sim$50
isolated, non-saturated stars), cross-match them by rejecting outliers,
determine the astrometric mapping between the images, and transform
them to the same reference system. 
Visual inspection of the registered frames show i) the prominent
Southward proper motion of \hds\ (in harmony with the Hipparcos
values), ii) a much fainter co-moving companion that we identify as
\hdsB\ (for details, see \figr{hd189mon}). The companion is clearly
separated from \hds\ on the 2MASS J, H and K scans (\figr{hd189mon},
right panel), and is listed in the 2MASS point source catalogue as
\hdsBm. Although not illustrated on \figr{hd189mon}, the
co-movement is also demonstrated by the TopHAT and FLWO 1.2m I-band
frames. No other co-moving companion is detected on these frames.



To quantify the proper motion of \hdsB, we have carried out astrometry
on the QuickV, POSS-II, 2MASS, TopHAT and FLWO1.2m observations.  We
used the 2MASS catalogue as astrometric reference, where the quoted
position uncertainty of bright, isolated sources is 120mas. We used
the previously established pixel centers of $\sim 50$ un-saturated
stars that we found by fitting Gaussian profiles. By running our
\fihat/\fistar\ star-finder algorithm on artificially
generated frames, we found typical errors of the centroid positions to
be on the order of 0.05pix (corresponding to 0.08\arcsec\ on QuickV,
0.05\arcsec\ on POSS-II and 2MASS, and 0.1\arcsec\ on TopHAT and
0.07\arcsec\ on FLWO1.2m). 
We then derived the second order astrometric mappings between the X,Y 
coordinates and the 2MASS astrometric reference 
(\hds(B) were omitted from the fit), 
and used this to transform the pixel coordinates of the frames
to the ICRS \citep[$\alpha$,
$\delta$; ][]{seidelmann02} system used by 2MASS. 
The rms around the
fit was $\sim0.2\arcsec$ for QuickV, $\sim 0.18\arcsec$ for POSS-II,
$\sim 0.05\arcsec$ for 2MASS, $\sim 0.3\arcsec$ for TopHAT, and
$0.1\arcsec$ for the FLWO1.2m, respectively.

\begin{figure}[h]
\plotone{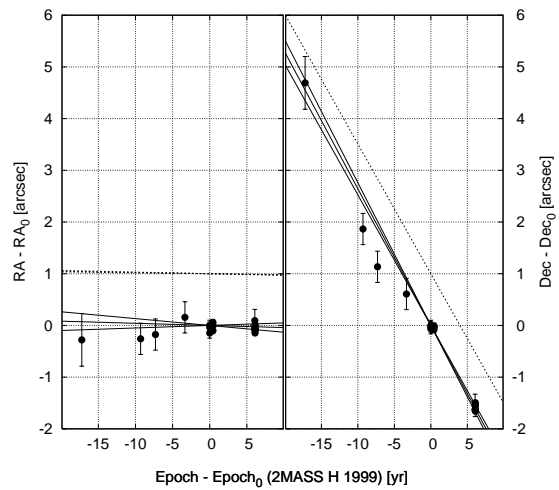}
\caption{
Proper motion of \hdsB\ in RA (left) and Dec (right) relative to the 2MASS
1999 position. The two panels are on the same scale. The central solid lines
show the linear fit to the data; the two other solid lines show the
same fits using slope and intersection parameters differing by $\pm 1 \sigma$.  
For reference, the dashed line shows the relative proper motion of \hds\
from Hipparcos, offset by 1.0 arcsec for clarity.
\label{fig:astrom}}
\end{figure}

The
derived proper motion of \hdsB\ is $\mu_{\alpha,B} = -4.1\pm9$ \masy,
$\mu_{\delta,B} = -264\pm12$ \masy (see \figr{astrom}), 
which are within $2\sigma$
of the Hipparcos proper motion of
\hds\ itself. Therefore, it is clear that, within uncertainties, \hds\ and
\hdsB\ share a common proper motion and so are likely to
comprise a bound system.  



\notetoeditor{This is the original location where Fig.2 was placed in
the latex file. If possible, please place Fig.2 on page 2}

\subsection{Radial Velocity}
\label{sec:radvel}

A common radial velocity is a further indication that the two stars are
physically associated. 
In order to test this possibility, we obtained spectroscopic
observations of both the primary star and the suspected
companion, using the Center for Astrophysics (CfA) Digital Speedometer
\citep[DS; ][]{latham92} at the 1.5m Tillinghast telescope of FLWO,
Arizona. 

Seven DS observations have been made of the star
\hds\, dating back to 1995, with the two most recent being
2005 Dec 10 and Dec 17. For each of these a radial velocity was
obtained on the CfA Native System velocity reference
\citep{stefanik99}. The mean and standard deviation of these 
measurements is 
$V_{\mathrm{rad}}$ = $-2.38 \pm 0.20$ \kms, 
with no significant evidence for a long-term velocity variation over the past 10
years. For \hdsB, two DS observations were made, on
2005 Dec 10 and 17, yielding a mean radial velocity
$V_{\mathrm{rad,B}} =-3.1 \pm 1.0$ \kms.
Thus, within observational uncertainties the difference in measured
velocities is consistent with the two stars being physical companions.

\subsection{Characteristics of the Star \hdsB}
\label{sec:char}

There is still a remote chance that the apparent companion could be a
distant giant star with high tangential velocity or a very close-by low
luminosity and velocity sub-dwarf, that happens to have the same sky position, 
radial velocity, and proper motion.

While the DS has the primary goal of radial-velocity measurements,
correlation of the spectrum with template spectra based on the
\citet{kurucz93} models also yields information on \teff, \logg\ and 
stellar rotation \vsini. We obtained $\logg = 4.5\pm0.3$ \--- a 
value appropriate to a main sequence M dwarf.  Correlation with
observed spectra of M dwarfs with spectral type ranging from M0.0 to
M5.5 gives best correlation near M3.5. We also observed \hdsB\ with
the FAST spectrograph on the FLWO 1.5m telescope.  Visual comparison of the
resulting spectra to known M dwarf spectra suggests a spectral type of
M4V. 
Based on its spectral type, plus apparent magnitude, we then infer a
distance to \hdsB\ consistent with the 19.3 pc distance to \hds\ .

\notetoeditor{This is the original location where Fig.3 was placed in
the latex file. If possible, please place Fig.3 on page 3}
\begin{figure}[h]
\plotone{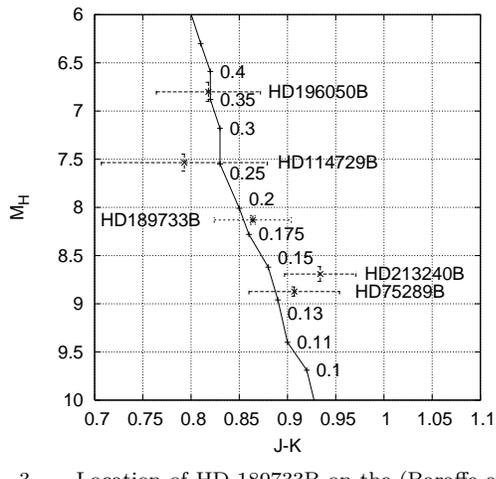}
\caption{
Location of \hdsB\ on the \citep{baraffe98} 5Gyr isochrone
with other M-dwarf binary companions of stars with known planets,
plotted from \citet{eggenberger04b}.
Stellar masses are labeled along the isochrone in units of \msun.
\label{fig:iso}}
\end{figure}


We independently estimated the spectral type of \hdsB\ from 2MASS
photometry. Because of the slightly overlapping profile of the
close-by, bright \hds\, we performed aperture photometry of \hdsB\ on
the 1999 2MASS scans after subtracting off the Gaussian profile of
\hds, and using $\sim 50$ isolated stars with original 2MASS
photometry as reference.  This analysis yields $\rm J=10.147\pm0.02$,
$\rm H=9.551\pm0.03$, and $\Ks=9.318\pm 0.02$.  These values are
within $0.03$mag of the 2MASS Point Source Catalogue values, but we
find smaller errors, and slightly redder $\rm J-K$ color index. We
transformed our measured $\rm J-\Ks$ value (namely $0.829 \pm 0.03$),
to $\rm J-K$ on the Bessell and Brett (BB) system following
\citet{carpenter05}, to obtain $\rm (J-K)_{BB}=0.864 \pm 0.04$.  Then
we derived the absolute magnitude $M_H$ ($8.13\pm0.045$) from our
measured H magnitude and an assumed distance equal to that of \hds\, 
and plotted these values on the color-magnitude diagram of
\citet{mugrauer05a}, which also plots a 5-Gyr isochrone from
\citet{baraffe98}. The position on the color magnitude diagram
(\figr{iso}) corresponds to a stellar mass of 0.175 to 0.2\msun,
which according to 
\citet{cox00} corresponds to an M
dwarf with spectral type of about M5. The good fit to the
isochrone supports our assumption that the distance to \hdsB\ is
similar to that to \hds.

From the above analyses we conclude that \hdsB\ is an M dwarf with
spectral type in the range M3.5 to M5, with common proper motion,
radial velocity, and distance to \hds. 
The relative positions (from astrometry, and assuming equidistance),
and relative velocities of the two stars (from proper motion and radial
velocity data), along with the 1\msun\ total system mass, are
consistent within 2$\sigma$ with the system being gravitationally
bound.
Although it would formally be
possible for an interloper M dwarf star to be passing very close to
\hds\ at the current epoch, with a relative speed so small as to make
it almost
gravitationally bound, this possibility is so
remote that we conclude that the two stars do indeed form a bound
system with projected separation about 216 AU.

\section{Orbital Characteristics of the Secondary}
\label{sec:orb}

If the true separation of \hds\ and \hdsB\ is close to the projected
separation, and the orbit is circular, and the total system mass is
$\sim$1\msun\ (0.82\msun\ for \hds\ from \citetalias{bouchy05} and
0.2\msun\ for \hdsB\ from \secr{char}) then the orbital period is
$\sim$3200 years, corresponding to $\sim$2\kms\ orbital motion.
For a face on orbit this would yield an observable $22$ \masy\  
differential proper motion in addition to the co-movement.

\begin{figure}[h]
\plotone{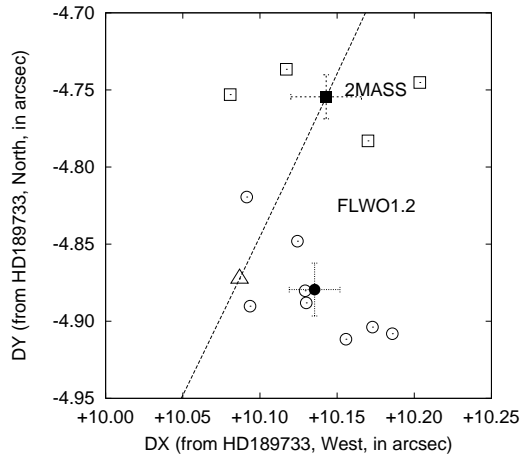}
\caption{
Differential proper motion of \hdsB\ relative to \hds. 
Open squares and circles indicate individual 2MASS (epoch 2000.073) 
and FLWO1.2 (epoch
2005.970) data points respectively. Filled square and circle represent
the mean of those respective data points, with error bars representing
1 standard deviation of those means. Dashed line depicts a circular
orbital path about \hds\ in the plane of the sky, passing through the
2MASS position. The triangle depicts the position at epoch 2005.973 for
such an orbit.
\label{fig:diffppm}}
\end{figure}

In an attempt to detect this, we used two pairs of 2MASS images in
the H and K bands obtained in 1999 and 2000 (one of these is the
right-hand image in \figr{hd189mon}), with mean epoch 2000.073, and
compared them with the 8 high resolution I-band frames of the field
taken with the FLWO 1.2m telescope at epoch 2005.970. We note that
although the other data were useful in confirmation of the common
proper motion (as shown in \secr{comprop}), because of the saturated
\hds\ image (POSS, Quick-V) and low S/N (TopHAT) they were not used in
this precise astrometry aimed at refining the differential proper motion.
The astrometry of \hds\ over the 5.897 year baseline yields a proper
motion of
$\mu_{\alpha} = -8 \pm 5$ \masy\ and 
$\mu_{\delta} = -245 \pm 8$ \masy.
Because this agrees with the more precise Hipparcos value within uncertainties
(\secr{comprop}), we adopt that value for the proper motion of \hds. 
For \hdsB\ we find 
$\mu_{\alpha,B} = -3 \pm 5$ \masy\ and 
$\mu_{\delta,B} = -272 \pm 5$ \masy, 
which imply a differential proper motion relative to \hds\ of 
$\Delta \mu_{\alpha}=   -1 \pm 5$ \masy\ and
$\Delta \mu_{\delta}=-21.2 \pm 5$ \masy
(\figr{diffppm}). 

Formally, the data indicate a detection of relative
proper motion at the 4$\sigma$ level. The position of the
companion, and its direction and magnitude of relative motion, 
are consistent with orbital motion in a clockwise
orbit roughly in the plane of the sky.

However, there  could be underlying systematic effects due
to the different instruments and bandpasses used for the earlier-epoch
2MASS data and the later-epoch FLWO1.2 data.
We estimated one of these, namely the effect of stellar profile merging
of \hdsB\ with \hds\ that is different on the 2MASS and FLWO1.2m
frames. By subtracting off the Gaussian profile of the primary, we
found that the derived proper motion of \hdsB\ changed by only 2\masy.
Nevertheless, while the detection of orbital motion 
roughly in the plane of the sky 
seems secure, it is
premature to derive specific orbital parameters for the system.

\notetoeditor{This is the original location where Fig.4 was placed in
the latex file. If possible, please place Fig.4 on page 4 (or 3)}




\section{Discussion}
\label{sec:conc}

Several studies have been done on the characteristics of close-in
planets orbiting the primary star of a multiple star system. Thus
\citet[][and references therein]{eggenberger04b}
found a tendency for massive planets (\msini\ greater than 2\mjup) to occur
preferentially in multiple-star systems.
However, \hdsb\ is a low-mass planet (1.15\mjup without $\sin i$
ambiguity), and yet it is in a multiple stellar system, thus weakening
the distinction between single and multiple stars. 

\hdsb\ exceeds all other known extrasolar planets in binary
systems in its proximity to its parent star ($a_{pl}$=0.031
AU). However, a number of planets in multiple-star systems are nearly as
close (e.g.~{\mbox Tau Boo b}, $a_{pl}$=0.05 AU; 
{\mbox HD 75289b}, $a_{pl}$=0.046 AU).

The data suggest (see \secr{orb}) that the binary orbit is likely to be
nearly face on, i.e.~the orbital plane would be nearly orthogonal
to the orbital plane of the planet \hdsb, which by virtue of its
transit we know to have an inclination of nearly $90^{\circ}$.
A detailed calculation based on the relative velocities and positions
of the two stars shows that the orbit of \hdsB\ and \hdsb\ cannot be
coplanar at the 4-$\sigma$ level (P\'al et al, in preparation).
When additional transiting planets are discovered in multiple star
systems, it should be possible to study the relation of the system
architecture (e.g.~mass, semi-major axis of the stellar secondary) to
planet properties in better detail.

Because the \hds\ system is so close-by, it should be possible to do
excellent astrometry over the next few years with modern high
precision astrometric techniques from ground or space.  The changing
radial velocity signal of either or both of \hds\ and \hdsB\ might
also be detectable after a few years of monitoring with current
high-precision radial velocity devices. High precision proper motion
and radial velocity measurements  should lead to a good
characterization of the orbital characteristics of the binary system
and their relation to the orbit of the transiting planet.  

Finally, we note that because this system is only 19 pc
from Earth and hence both stellar components are unusually bright for
their spectral types, many additional follow-on
observations requiring high resolution spectroscopy or high precision
photometry will be feasible. This should permit a full characterization
of the system including the possible detection of additional low mass
components.

\acknowledgments
This work was funded by NASA grant NNG04GN74G. 
Work by G.~\'A.~B.\ was supported by NASA through grant
HST-HF-01170.01-A Hubble Fellowship.
A.~P.~wishes to acknowledge the hospitality of the Harvard-Smithsonian
Center for Astrophysics, where part of this work has been carried out.
Work of A.~P.~was also supported by Hungarian OTKA grant T-038437. 
D.~W.~L. and R.~P.~S thank the Kepler mission for support through NASA
Cooperative Agreement NCC2-1390. We acknowledge the use of the 2MASS
survey frames, and the Palomar Sky Survey digital scans. We thank Perry
Berlind at FLWO for taking the FAST spectra of \hdsB, and Emilio Falco
for the follow-up exposures using the 1.2m telescope.



\end{document}